\documentclass[superscriptaddress,twocolumn,aps]{revtex4}
\usepackage{graphicx}
\usepackage{amsmath}
\begin{document}
\title{Instability of a Thin Conducting Foil Accelerated by a Finite Wavelength Intense Laser}
\author{B. Eliasson}
\affiliation{SUPA, Physics Department,
John Anderson Building,
Strathclyde University,
Glasgow G4 0NG,
Scotland, UK.}
\email{bengt.eliasson@strath.ac.uk}
\begin{abstract}
We derive a theoretical model for the Rayleigh-Taylor (RT)-like instability for a thin foil accelerated
by an intense laser, taking into account finite wavelength effects in the laser wave field. The latter
leads to the diffraction of the electromagnetic wave off the periodic
structures  arising from the instability of the foil, which significantly modifies
the growth rate of the RT-like instability when the perturbations on the foil
have wavenumbers comparable to or larger than the laser wavenumber. In particular,
the growth rate has a local maximum at a perturbation wavenumber approximately equal to the laser wavenumber.
The standard RT instability, arising from a pressure difference between the two sides of a foil,
is approximately recovered for perturbation wavenumbers
smaller than the laser wavenumber. Differences in the results for circular and linear polarization
of the laser light are pointed out.
The model has significance to radiation pressure acceleration
of thin foils and to laser-driven inertial confinement fusion schemes, where RT-like instabilities
are significant obstacles.
\end{abstract}
\date{23 July 2014}
\maketitle
The Rayleigh-Taylor (RT) instability, or RT-like instabilities \cite{Ott}, is one of the main
obstacles preventing a greater success of the radiation pressure acceleration
scheme for accelerating thin foils of ions by
intense lasers \cite{Pegoraro,Yan08,Chen,Robinson08,Liu11,Palmer12,Adusumilli12}, and in the realization of inertial
confinement fusion via laser compression and
heating of fuel pellets \cite{Takabe}.
While the RT instability was originally associated with a heavier fluid on top
of a lighter fluid in a gravitational field \cite{Taylor},
similar instabilities occur for plasmas confined by magnetic fields (e.g.~Ref. \cite{Goldston}),
and when a thin foil is accelerated by the pressure difference
between the two sides of the foil \cite{Ott,Pegoraro}.
The growth rate of the RT instability for laser accelerated plasma is
typically proportional to $\sqrt{gk}$, where $g$ is the
acceleration and $k$ the wavenumber of the surface perturbation. This predicts
that the instability grows indefinitely for large wavenumbers;
while in some experiment and simulations,
the RT instability gives rise to structures with a spatial periodicity
comparable to the laser wavelength \cite{Palmer12}.
The assumption of a constant normal pressure force is reasonable as long
as the perturbations of the foil are relatively small
and when the length-scales of the perturbations are much larger than the wavelength of the laser \cite{Pegoraro}.
However, the laser light has a finite wavelength, and
is scattered off the periodic structures leading to a diffraction of the
electromagnetic (EM) wave.
Therefore the directions of the scattered light will be quantized, and
the "pressure" picture can only be expected to be approximate for
monochromatic laser light.
 Theoretical investigations of the instabilities resulting from
 the scattering of EM waves off plasma surface perturbations
 include the RT instability of an over-dense
plasma layer \cite{Gamaly93} using a magnetohydrodynamic-like model for the
plasma, and the scattering off surface plasma waves \cite{Macchi02} where
the electron dynamics is the dominant source of the instability.
The aim of this Letter is to solve the
scattering problem and to derive a model for the instability of an ultra-thin,
perfectly conducting foil accelerated by the radiation pressure of
a finite wavelength intense laser.

We assume that the laser interacts with a foil where
the electron density is much higher than the critical density
so that no laser light penetrates the foil.
We carry out the calculations in a frame moving with the velocity of the unperturbed foil.
In this frame, the dynamics of the initially small-amplitude perturbations of the foil is non-relativistic.
The results obtained in the moving frame can later be Lorentz transformed to the laboratory frame, but
we will here for simplicity assume that the speed of the foil is non-relativistic.
The velocity ${\bf v}$ of the foil relative to the accelerated frame is governed by the momentum equation
\begin{equation}
  M\bigg(\frac{\partial}{\partial t}+v_x\frac{\partial}{\partial x}+v_y\frac{\partial}{\partial y}\bigg){\bf v}={\bf F}-M g_{0}\widehat{\bf z},
  \label{Eq_motion}
\end{equation}
where $M(x,y,t)$ is the surface mass density, $g_0=F_0/M_0$ is the acceleration of the unperturbed foil in the $z$-direction,
$M_0$ is the unperturbed areal mass density of the foil,
$F_{0}=2I_0/c$ is the radiation pressure force,
$I_0$ is the incident laser intensity, and $c$ the speed of light in vacuum.
The force ${\bf F}$ is due to the space- and time-dependent EM field acting on the foil.
For an unperturbed foil, with $M=M_0$, the force ${\bf F}$ would be exactly canceled by the
inertial force $-M_0 g_{0}\widehat{\bf z}$, but
due to perturbations in the foil, the forces are not exactly canceled, which will lead to the RT-like instability.
The mass density is governed by the continuity equation
\begin{equation}
  \frac{\partial M}{\partial t}+\frac{\partial (M v_x)}{\partial x}+\frac{\partial (M v_y)}{\partial y}=0.
\end{equation}
The foil surface can be parameterized as $S(x,y,z,t)=z-\eta(x,y,t)=0$,
where $\eta$ is the surface elevation of the foil in the $z$-direction.
The velocity and surface elevation are connected through
the kinematic condition
\begin{equation}
  \frac{\partial\eta}{\partial t}={\bf v}\cdot\nabla S.
  \label{eq_kin}
\end{equation}
Equations (\ref{Eq_motion})--(\ref{eq_kin}) are completed by initial conditions on $\eta$ and on $M$ and ${\bf v}$ at $z=\eta$.

First we notice that the assumption of a constant radiation pressure force $F_0$
acting perpendicularly to the surface on one side of the foil \cite{Ott,Pegoraro}
would lead to that ${\bf F}=F_0 \nabla S$ in Eq. (\ref{Eq_motion}) and to a ``standard'' RT instability with
 the growth rate $\sqrt{g_0 k}$.
Here we will instead determine ${\bf F}$ by taking into
account that the electric and magnetic fields
${\bf E}$ and ${\bf B}$ evolve in time according to Maxwell's equations,
obeying boundary conditions at the foil surface as well as radiating boundary conditions
far away from the foil.
We assume that the foil is perfectly
conducting, and therefore the electric field parallel to the
surface and the magnetic field perpendicular to the surface are zero
in a system (denoted by primed variables) moving with the same velocity as the surface,
with the boundary conditions expressed as ${\bf E}'\times\nabla S=0$ and ${\bf B}'\cdot \nabla S=0$ at $z=\eta$.
Assuming non-relativistic velocities in the moving frame, the magnetic and electric fields are Galilei transformed
from the system moving with the foil surface (primed variables) to the accelerated frame (unprimed variables) as
${\bf E}'={\bf E}+{\bf v}\times {\bf B}$ and ${\bf B}'={\bf B}-{\bf v}\times {\bf E}/c^2\approx {\bf B}$. (The
term $-{\bf v}\times {\bf E}/c^2$ will only contribute to the boundary conditions with terms of order $v^2/c^2$ compared to unity, and is therefore neglected.) This gives
\begin{equation}
  {\bf B}\cdot \nabla S=0
  \label{B_boundary}
\end{equation}
for the magnetic field, while for the electric field we have
$0={\bf E}'\times\nabla S=({\bf E}+{\bf v}\times {\bf B})\times\nabla S={\bf E}\times\nabla S
-{\bf v}({\bf B}\cdot\nabla S)+ {\bf B}({\bf v}\cdot\nabla S)$, where ${\bf B}\cdot\nabla S=0$ and
${\bf v}\cdot\nabla S=\partial\eta/\partial t$, giving
\begin{equation}
 {\bf E}\times\nabla S+{\bf B}\frac{\partial \eta}{\partial t}=0
 \label{E_boundary}
\end{equation}
at $z=\eta$.

The force acting on the surface can be calculated
using the EM volume force \cite{Jackson}
\begin{equation}
  {\bf f}=\nabla\cdot{\bar{\bar{\boldsymbol{\sigma}}}}-\epsilon_0\frac{\partial}{\partial t}{\bf E}\times{\bf B},
\end{equation}
where
\begin{equation}
  \sigma_{ij}=\epsilon_0\bigg(E_i E_j-\frac{1}{2}\delta_{ij} E^2 \bigg)
  + \frac{1}{\mu_0}\bigg(B_i B_j-\frac{1}{2}\delta_{ij} B^2 \bigg)
\end{equation}
is the Maxwell stress tensor on component form, $\delta_{ij}$ represents the unit tensor, $\epsilon_0$ is
the electric permittivity in vacuum, and $\mu_0=1/(\epsilon_0 c^2)$ is the magnetic permeability in vacuum.
Integrating ${\bf f}$ from $z=\eta-\varepsilon$ to $\eta+\varepsilon$,
assuming that ${\bf E}$ and ${\bf B}$ are zero for $z>\eta$, and letting
$\varepsilon\rightarrow 0$ gives the EM area force
  ${\bf F}=-\bar{\bar{\boldsymbol{\sigma}}}\cdot\nabla S-\epsilon_0{\bf E}\times{\bf B}{\partial\eta}/{\partial t}$,
which, using the boundary conditions (\ref{B_boundary}) and (\ref{E_boundary}), simplifies to
\begin{equation}
  {\bf F}=\frac{1}{2}\bigg(\frac{B^2}{\mu_0}-\epsilon_0 E^2 \bigg)\nabla S.
  \label{Eq_F2}
\end{equation}
It should be emphasized that in Eq.~(\ref{Eq_F2}), ${\bf E}$ and ${\bf B}$ are the
total electric and magnetic fields at the foil surface, to be determined below.

Perturbing and linearizing the system of equations
(\ref{Eq_motion})--(\ref{eq_kin}) and (\ref{Eq_F2}) around the equilibrium solution ${\bf v}=0$, $\eta=0$, $M=M_0$, $S=z$,
${\bf E}={\bf E}_0(t)$, and ${\bf B}={\bf B}_0(t)$, gives
\begin{equation}
  \frac{\partial^4 \eta_1}{\partial t^4}+\frac{F_0^2}{M_0^2}\bigg(\frac{\partial^2\eta_1}{\partial x^2}+\frac{\partial^2\eta_1}{\partial y^2}\bigg)=\frac{1}{M_0}\frac{\partial^2}{\partial t^2}\bigg(\frac{{\bf B}_0\cdot{\bf B}_1}{\mu_0}-\epsilon_0{\bf E}_0\cdot{\bf E}_1\bigg),
  \label{eq29}
\end{equation}
where the subscript 1 denotes small-amplitude, first-order perturbations.
For circularly polarized light, the zeroth order EM force is
\begin{equation}
  F_0=\frac{1}{2}\bigg(\frac{B_0^2}{\mu_0}-\epsilon_0E_0^2\bigg),
  \label{eq29c}
\end{equation}
while for linearly polarized light
a time averaging over one laser period removes second harmonics and reduces $F_0$
a factor 2 for given amplitudes $B_0$ and $E_0$.
Equation (\ref{eq29}) is completed by finding the dependence of ${\bf E}_1$ and ${\bf B}_1$
on $\eta_1$. The general form of Eq.~(\ref{eq29}) is that of a mode-coupling equation,
where the low-frequency perturbations of the foil are driven by
the coupling (beating) between the large amplitude EM wave (${\bf B}_0$, ${\bf E}_0$) and its
small-amplitude side-bands (${\bf B}_1$, ${\bf E}_1$).

Writing out the components of the boundary conditions (\ref{B_boundary}) and (\ref{E_boundary})
gives
\begin{equation}
  {B}_{z}-{B}_{x}\frac{\partial \eta}{\partial x}-{B}_{y}\frac{\partial \eta}{\partial y}=0,
  \label{eq_Brz2}
\end{equation}
\begin{equation}
  {E}_{y}+{E}_{z}\frac{\partial \eta}{\partial y}+{B}_{x}\frac{\partial \eta}{\partial t}=0,
  \label{eq_Ery2}
\end{equation}
and
\begin{equation}
  {E}_{x}+{E}_{z}\frac{\partial \eta}{\partial x}-{B}_{y}\frac{\partial \eta}{\partial t}=0,
  \label{eq_Erx2}
\end{equation}
at $z=\eta$. An incident EM wave will be reflected by the foil, and perturbations
in the foil surface will lead to the refraction of the wave.
The electric and magnetic fields can be written ${\bf E}={\bf E}_{i0}+{\bf E}_{r}$ and
${\bf B}={\bf B}_{i0}+{\bf B}_{r}$, where ${\bf E}_{i0}$ and ${\bf B}_{i0}$ are the
fields of the incident wave and ${\bf E}_{r}$ and ${\bf B}_{r}$ of the reflected wave.
In what follows, we will show details of the calculations for a circularly polarized
incident wave, and at the end only state the final result also for a linearly polarized wave.
More details of the derivations will be given elsewhere.
For an incident, right-hand circularly polarized EM wave propagating in the $z$-direction, we have
\begin{equation}
  {\bf E}_{i0}=\frac{\widehat{\bf e}}{2} \widehat{E}_{i0} e^{i\theta_i}+\rm{c.c.}
  \label{eq_Ei1}
\end{equation}
and
\begin{equation}
  {\bf B}_{i0}=\frac{\widehat{\bf e}}{2} \widehat{B}_{i0} e^{i\theta_i}+\rm{c.c.},
  \label{eq_Bi1}
\end{equation}
where $\widehat{\bf e}=\widehat{\bf x}+i\widehat{\bf y}$ describes the polarization,  $\widehat{\bf x}$ and $\widehat{\bf y}$ are unit vectors in the $x$- and $y$-direction, $\theta_i=k_0 z-\omega_0 t$ is the phase of the
incident wave, $k_0$ is the incident wave-number, $\omega_0=ck_0$ the frequency, and $\widehat{E}_{i0}=ic \widehat{B}_{i0}$.
For linearly polarized light with the electric field along the $x$-axis, we would instead have ${\bf E}_{i0}=(\widehat{\bf x}/2)\widehat{E}_{i0}\exp(i\theta_i)+$c.c.,
${\bf B}_{i0}=(\widehat{\bf y}/2)\widehat{B}_{i0}\exp(i\theta_i)+$c.c., and $\widehat{E}_{i0}=c\widehat{B}_{i0}$.
We next assume small perturbations of the surface, so that $\eta(x,y,t)=\eta_1(x,y,t)$, where
$|\nabla| |\eta_1| \ll 1$. (It implies small wave steepness $|\nabla\eta_1| \ll 1$
and that $|\eta_1 \nabla| \ll 1$ when acting on ${\bf E}$ and ${\bf B}$.)
Then ${\bf E}_{z=\eta}\approx{\bf E}_{0, z=0}+{\bf E}_{1, z=0}+\eta_1(\partial{\bf E}_0/\partial z)_{z=0}$ and
${\bf B}_{z=\eta}\approx{\bf B}_{0, z=0}+{\bf B}_{1, z=0}+\eta_1(\partial{\bf B}_0/\partial z)_{z=0}$,
where $|{\bf E}_1|\ll |{\bf E}_0|$ and $|{\bf B}_1|\ll |{\bf B}_0|$.
At $z=0$, we have $\theta_i=\theta_0=-\omega_0 t$.
Writing ${\bf E}_r=\widetilde{\bf E}_r e^{i\theta_0}/2+$c.c.~and ${\bf B}_r=\widetilde{\bf B}_r e^{i\theta_0}/2+$c.c.,
and linearizing the boundary conditions (\ref{eq_Brz2})--(\ref{eq_Erx2}), we
have at $z=0$,
\begin{equation}
  \widetilde{B}_{rz1}-\widetilde{B}_{rx0}\frac{\partial \eta_1}{\partial x}-\widetilde{B}_{ry0}\frac{\partial \eta_1}{\partial y}
  =\widehat{B}_{i0}\bigg(\frac{\partial\eta_1}{\partial x}+i\frac{\partial\eta_1}{\partial y}\bigg),
  \label{eq_Brz6}
\end{equation}
\begin{equation}
  \widetilde{E}_{ry1}+\eta_1\frac{\partial \widetilde{E}_{ry0}}{\partial z}+\widetilde{B}_{rx0}\frac{\partial \eta_1}{\partial t}
  =\widehat{B}_{i0}\bigg(i\omega_0\eta_1-\frac{\partial \eta_1}{\partial t} \bigg),
  \label{eq_Ery6}
\end{equation}
and
\begin{equation}
  \widetilde{E}_{rx1}+\eta_1\frac{\partial \widetilde{E}_{rx0}}{\partial z}-\widetilde{B}_{ry0}\frac{\partial \eta_1}{\partial t}
  =-i\widehat{B}_{i0}\bigg(i\omega_0\eta_1-\frac{\partial \eta_1}{\partial t} \bigg).
  \label{eq_Erx6}
\end{equation}

To zeroth order, the boundary conditions at the foil surface $z=0$ is that the electric field parallel to the foil is zero,
${\bf E}_0=0$, and
therefore ${\bf E}_{r0}=-{\bf E}_{i0}$, and it follows from Maxwell's
equations that ${\bf B}_{r0}=+{\bf B}_{i0}$ at $z=0$.
Since ${\bf E}_0=0$ and ${\bf B}_0=2{\bf B}_{i0}$ in Eqs.~(\ref{eq29}) and (\ref{eq29c}),
it is apparent that the foil is
accelerated by the magnetic pressure of the EM field.
The unidirectional wave equations
  ${\partial {\bf E}_{r0}}/{\partial t}-c{\partial {\bf E}_{r0}}/{\partial z}=0$
and
${\partial {\bf B}_{r0}}/{\partial t}-c{\partial {\bf B}_{r0}}/{\partial z}=0$
of the reflected wave have the boundary conditions
${\bf E}_{r0}=-({\widehat{\bf e}}/{2})\widehat{E}_{i0}e^{i\theta_0(t)}+ \mbox{c.c.}$,
and
${\bf B}_{r0}=({\widehat{\bf e}}/{2})\widehat{B}_{i0}e^{i\theta_0(t)}+ \mbox{c.c.}$,
at $z=0$, with the solutions
${\bf E}_{r0}=-({\widehat{\bf e}}/{2})\widehat{E}_{i0}e^{i\theta_0(t')}+ \mbox{c.c.}$
and
${\bf B}_{r0}=({\widehat{\bf e}}/{2})\widehat{B}_{i0}e^{i\theta_0(t')}+ \mbox{c.c.}$,
where the retarded time $t'$
is obtained from $ct'=\xi$ with $\xi=z+ct$.
It follows that
$
\widetilde{\bf E}_{r0}=-\widehat{\bf e}\widehat{E}_{i0}e^{i\theta_0(t')-i\theta_0(t)}
$
and
$
\widetilde{\bf B}_{r0}=\widehat{\bf e}\widehat{B}_{i0}e^{i\theta_0(t')-i\theta_0(t)}.
$
Using that $\partial t'/\partial z=1/c$ and $\widehat{E}_{i0}=ic\widehat{B}_{i0}$, we have
$\widetilde{\bf E}_{r0}|_{z=0}=-i\widehat{\bf e} c\widehat{B}_{i0}$, $\widetilde{\bf B}_{r0}|_{z=0}=\widehat{\bf e} \widehat{B}_{i0}$,
$
  {\partial\widetilde{\bf E}_{r0}}/{\partial z}|_{z=0}=-\widehat{\bf e} \omega_0 \widehat{B}_{i0},
$
and
$
  {\partial\widetilde{\bf B}_{r0}}/{\partial z}|_{z=0}=-i\widehat{\bf e} k_0 \widehat{B}_{i0}.
$, which is used in Eqs.~(\ref{eq_Brz6})--(\ref{eq_Erx6}).

We assume a 4-wave model in which the EM wave is scattered
into two EM sidebands off the ripples in the foil surface, so that
$\widetilde{\bf E}_{r1}=\widehat{\bf E}_{r1+}\exp(-i\omega t+ik_xx+ik_yy+ik_{z+}z)+\widehat{\bf E}_{r1-}^*\exp(i\omega^* t-ik_xx-ik_yy-ik_{z-}^*z)$, $\widetilde{\bf B}_{r1}=\widehat{\bf B}_{r1+}\exp(-i\omega t+ik_xx+ik_yy+ik_{z+}z)+\widehat{\bf B}_{r1-}^*\exp(i\omega^* t-ik_xx-ik_yy-ik_{z-}^*z)$, and $\eta_1=\widehat{\eta}_1\exp(-i\omega t+ik_xx+ik_yy)+\widehat{\eta}_1^*\exp(i\omega^* t-ik_xx-ik_yy)$.
The vacuum wave equations for the scattered light, ${\partial^2 {\bf E}_{r1}}/{\partial t^2}-c^2\nabla^2 {\bf E}_{r1}=0$
and ${\partial^2 {\bf B}_{r1}}/{\partial t^2}-c^2\nabla^2 {\bf B}_{r1}=0$, then give the dispersion relation
\begin{equation}
  (\pm\omega+ck_0)^2-c^2(k_\perp^2+k_{z\pm}^2)=0,
  \label{disp_r}
\end{equation}
where $k_\perp^2=k_x^2+k_y^2$.
Equation (\ref{disp_r}) has the solutions $k_{z\pm}=\mp\sqrt{(k_0\pm \omega/c)^2-k_\perp^2}$, where
the branches of the square root are chosen such that ${\rm imag}(k_{z\pm})<0$ for ${\rm imag}(\omega)>0$.
This gives radiating boundary conditions with waves propagating out from the foil and vanishing at $z=-\infty$, which is consistent with the model.
For $k_0>k_\perp$, the scattered wave is diffracted and propagates out from the foil at an
angle $\varphi$ to the negative $z$-axis,
given by $\sin\varphi\approx k_\perp/k_0$, while for $k_0<k_\perp$ the scattered wave is evanescent
and decays rapidly with the distance from the foil. Separating wave modes proportional to $\exp(-i\omega t+ik_xx+ik_yy)$
and $\exp(i\omega^* t-ik_xx-ik_yy)$, the boundary conditions (\ref{eq_Brz6})--(\ref{eq_Erx6}) yield
the Fourier coefficients
    $\widehat{B}_{rz1+}=2\widehat{B}_{i0}(ik_x-k_y)\widehat{\eta}_1$,
    $\widehat{B}_{rz1-}=2\widehat{B}_{i0}^*(ik_x+k_y)\widehat{\eta}_1$,
  $\widehat{E}_{ry1+}
  =2i\widehat{B}_{i0}(\omega_0+\omega)\widehat{\eta}_1$,
  $\widehat{E}_{ry1-}
  =-2i\widehat{B}_{i0}^*(\omega_0-\omega)\widehat{\eta}_1$,
  $\widehat{E}_{rx1+}
  =2\widehat{B}_{i0}(\omega_0+\omega)\widehat{\eta}_1$,
and
  $\widehat{E}_{rx1-}
  =2\widehat{B}_{i0}^*(\omega_0-\omega)\widehat{\eta}_1$.
From the divergence condition $\nabla\cdot{\bf E}_r=0$ to the left of the foil, we obtain
  $\widehat{E}_{rz1+}=-2\widehat{B}_{i0}(\omega_0+\omega)(k_x+ik_y)\widehat{\eta}_1/k_{z+}$
and
  $\widehat{E}_{rz1-}=-2\widehat{B}_{i0}^*(\omega_0-\omega)(k_x-ik_y)\widehat{\eta}_1/k_{z-}$,
and from the $x$- and $y$-components of Faraday's law $\partial {\bf B}_r/\partial t=-\nabla\times{\bf E}_r$,
we have
  $\widehat{B}_{rx1+} = -2i\widehat{B}_{i0}(k_y^2+k_{z+}^2-i k_x k_y)\widehat{\eta}_1/k_{z+}$,
  $\widehat{B}_{ry1+} = 2\widehat{B}_{i0}(k_x^2+k_{z+}^2+i k_x k_y)\widehat{\eta}_1/k_{z+}$,
  $\widehat{B}_{rx1-} = -2i\widehat{B}_{i0}^*(k_y^2+k_{z-}^2+i k_x k_y)\widehat{\eta}_1/k_{z-}$,
and
  $\widehat{B}_{ry1-} = -2\widehat{B}_{i0}^*(k_x^2+k_{z-}^2-i k_x k_y)\widehat{\eta}_1/k_{z-}$.

We next insert these results into Eq.~(\ref{eq29}) and separate terms proportional to $\exp(-i\omega t+ik_xx+ik_yy)$ and/or
$\exp(i\omega^* t-ik_xx-ik_yy)$.  This gives the dispersion relation for the RT-like instability for
circularly polarized incident laser light,
\begin{equation}
   \omega^4-g_0^2k_\perp^2=i \frac{\omega^2 g_0}{2}\sum_{+,-}\frac{k_\perp^2+2k_{z\pm}^2}{k_{z\pm}},
   \label{disp1}
\end{equation}
where $k_{z\pm}$ is given by the solutions of Eq.~(\ref{disp_r}), and $g_0=F_0/M_0$.
An analogous calculation for linearly polarized light with ${\bf E}_{i0}=(\widehat{\bf x}/2)\widehat{E}_{i0}\exp(i\theta_i)+$c.c.,
${\bf B}_{i0}=(\widehat{\bf y}/2)\widehat{B}_{i0}\exp(i\theta_i)+$c.c., and $\widehat{E}_{i0}=c\widehat{B}_{i0}$ yields the dispersion relation
\begin{equation}
   \omega^4-g_0^2k_\perp^2=i \omega^2 g_0\sum_{+,-}\frac{k_x^2+k_{z\pm}^2}{k_{z\pm}}.
   \label{disp1L}
\end{equation}
The dispersion relations (\ref{disp1}) and (\ref{disp1L}) have one positive imaginary root $\omega=i\omega_I$, which
gives rise to a purely growing instability with growth rate $\omega_I$.
If the right-hand sides of Eqs.~(\ref{disp1}) and (\ref{disp1L}) are neglected, then we recover
the standard RT instability with the growth-rate
$\omega_I=\sqrt{g_0k_\perp}$. There also exist two real-valued roots which
give rise to oscillatory solutions, similarly as for the standard RT instability \cite{Ott}.
To compare with experiments and simulations, we notice first that a critical
dimensionless parameter of the system is the normalized acceleration $g_0/(c^2k_0)$, which
can be expressed in terms of commonly used laser-plasma parameters as
$g_0/(c^2k_0)=2\sigma (Z_i m_e/m_i)(n_{cr}/n_e)a_0^2/(k_0 d)$,
where $Z_i$ is the charge state of the ions, $m_e$ and $m_i$ the electron and ion mass, $n_e/n_{cr}$ is the ratio of the electron density to the critical density, $a_0=eE_{i0}/(m_e c \omega_0)$ is the normalized laser amplitude, $d$ is the foil thickness, and the coefficient $\sigma=1/2$ for linearly polarized light and $\sigma=1$ for circularly polarized light.
For example, Yan et al. \cite{Yan08} used
circularly polarized light ($\sigma=1$) in their simulations to study the radiation
pressure acceleration of a proton $H^+$ foil ($Z_i=1$, $m_i = 1836 m_e$)
with $n_0/n_{cr}=10$, $k_0 d=0.63$, and $a_0=5$, giving $g_0/(c^2k)\approx 4.3\times 10^{-3}$. On the other hand,
Palmer et al. \cite{Palmer12} used linearly
polarized light ($\sigma=1/2$) in their experimental and simulation study of the RT instability
of a carbon $C^{6+}$ foil ($Z_i=6$, $m_i\approx 12\times 1836 \times m_e$) with
$n_e/n_{cr}=10^3$, and $k_0 d=0.03$. Using their values $a_0=10$ and $a_0=20$ gives $g_0/(c^2k)\approx 9.1\times 10^{-4}$
and $3.6\times 10^{-3}$, respectively.

\begin{figure}
\centering
\includegraphics[width=8.5cm]{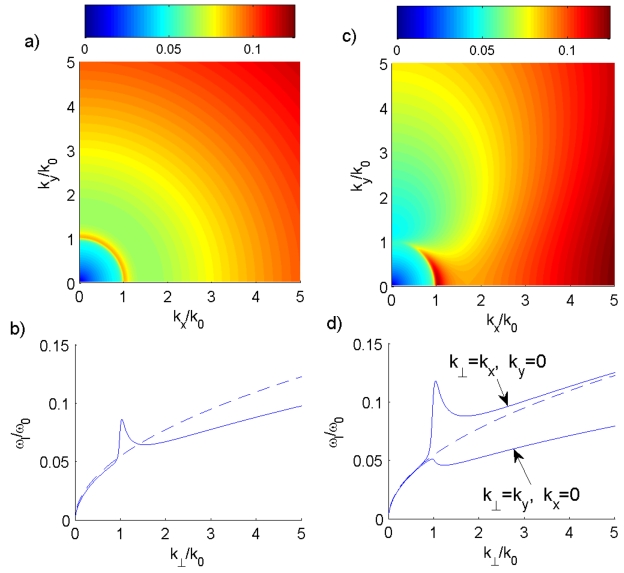}
\caption{The normalized growth rate $\omega_I/\omega_0$ of the instability for circularly polarized waves [panels a), b)] obtained
from Eq.~(\ref{disp1}), and
linearly polarized waves [panels c), d)] obtained from Eq. (\ref{disp1L}).  Top panels a), c) show
color plots of the growth rate as a function
of ($k_x/k_0$, $k_y/k_0$), while the lower panels b), d) show line plots of the growth rate (solid lines).
A comparison is made with the standard Rayleigh-Taylor instability (dashed line).}
\end{figure}

Figure 1 shows the growth rates of the instability for a typical value $g_0/(c^2k)= 3\times 10^{-3}$.
For the case of circularly polarized light, it is noticeable from Figs.~1a and 1b that
the growth rate of the instability is close to the one of the standard RT
instability for $k_\perp<k_0$, has a sharply peaked maximum at $k_\perp\approx k_0$,
and has a lower growth-rate than the standard RT instability for $k_\perp\gtrsim 1.5 k_0$.
For linearly polarized light, we see in Figs.~1c and 1d that the instability is strongly anisotropic,
with a larger growth rate for perturbation wavenumbers in the $x$-direction,
parallel to the electric field and perpendicular to the magnetic field of the
incident EM wave. Similar situations often occur in plasmas confined
by a non-oscillatory magnetic field and gives rise to RT-like instabilities,
such as the gravitational and flute instabilities \cite{Goldston}, where the
perturbation wavenumbers of the fastest growing unstable waves are at angles almost perpendicular
to the magnetic field. The RT-like instability has also a large growth rate
for $k_\perp\gg k_0$, where the instability can be
expected to saturate nonlinearly by forming small-scale structures but without disrupting the foil.
The most severe instability is at $k_\perp\approx k_0$, which leads to the
disruption of the foil and to the broadening of the energy spectrum \cite{Liu11}.
A scheme tailored to reduce the maximum of the growth rate at $k_\perp\approx k_0$ of the RT-like instability
could potentially make laser driven radiation pressure acceleration and compression schemes
more tractable.


{\bf Acknowledgments} Useful discussions with C.~S.~Liu, X.~Shao and T.~C.~Liu
at University of Maryland, and Z.-M.~Cheng, T.~Heelis, and A.~W.~Cross at University of Strathclyde
are gratefully acknowledged.

\end{document}